\documentclass[a4paper,fleqn]{cas-sc}

\usepackage[numbers]{natbib}
\usepackage[section]{placeins}

\def\tsc#1{\csdef{#1}{\textsc{\lowercase{#1}}\xspace}}
\tsc{WGM}
\tsc{QE}

\begin{document}
\let\WriteBookmarks\relax
\def\floatpagepagefraction{1}
\def\textpagefraction{.001}

\shorttitle{ML for Size-Dependent Temperature Scales}    

\shortauthors{Fan et al.}  

\title [mode = title]{Machine-learning extraction of size-dependent temperature scales in the 2D XY model}  



%

\author[1,2]{Qingao Fan}



\ead{fanqingao@stu.kust.edu.cn}


\credit{Conceptualization, Methodology, Software, Investigation, Writing – original draft}

\affiliation[1]{organization={Faculty of Science, Kunming University of Science and Technology},
            addressline={}, 
            city={Kunming},
            postcode={650500}, 
            state={Yunnan},
            country={China}}

\affiliation[2]{organization={Yunnan Key Laboratory of Complex Systems and Brain-Inspired Intelligence},
            addressline={}, 
            city={Kunming},
            postcode={650500}, 
            state={Yunnan},
            country={China}}

\author[1,2]{Xu Li}[orcid=0000-0001-8556-4558]

\cormark[1]


\ead{lixucn@kust.edu.cn}


\credit{Conceptualization, Methodology, Supervision, Writing – review \& editing}

\author[1,2]{Tingting Xue}[orcid=0000-0002-6098-4094]

\cormark[2]


\ead{tingtingxue@kust.edu.cn}


\credit{Conceptualization, Supervision, Writing – review \& editing}

\cortext[1]{Corresponding author}
\cortext[2]{Corresponding author}

\fntext[1]{This work was supported by the National Natural Science Foundation of China (Grant No.~12135003).}


\begin{abstract}
Machine learning has become a useful tool for studying phase transitions in statistical systems. For the two-dimensional classical XY model, however, the topological character of the Berezinskii-Kosterlitz-Thouless (BKT) transition and pronounced finite-size effects make it nontrivial to extract robust size-dependent pseudo-critical temperatures from configuration data. Existing studies often stop at phase classification, leaving open how standard neural-network outputs can be turned into quantitatively testable observables. Here we develop a machine-learning-assisted framework for the 2D XY model that uses standard network outputs to extract the size-dependent sequence of pseudo-critical temperatures $T^*(L)$. Specifically, we generate Monte Carlo configurations using embedded cluster updates, train a standard ResNet18 only on samples from the Quasi-ordered Phase and the Disordered Phase, and determine $T^*(L)$ from bootstrap-averaged probability curves using the 50\% crossing criterion. We then analyze the finite-size drift of this temperature sequence using BKT-motivated scaling and compare it with susceptibility-peak temperatures. The resulting temperature sequence shows a systematic finite-size drift consistent with BKT-type behavior and remains in the same fluctuation window as the susceptibility peak, supporting its interpretation as a finite-size pseudo-critical temperature. More broadly, this framework provides a practical route for converting standard neural-network outputs into physically interpretable finite-size observables in systems with strong crossover or topological transition signatures.
\end{abstract}


\begin{keywords}
machine learning \sep finite-size effects \sep BKT transition \sep 2D XY model \sep temperature scales \sep uncertainty quantification
\end{keywords}

\maketitle

\section{Introduction}\label{sec:intro}

The two-dimensional classical XY model is a paradigmatic system for studying collective behavior in low-dimensional systems with continuous symmetry, topological excitations, and BKT physics. In accordance with the absence of true long-range order in two-dimensional systems with continuous symmetry~\cite{mermin1966}, its low-temperature phase exhibits quasi-long-range order rather than conventional long-range order. The associated Berezinskii-Kosterlitz-Thouless (BKT) transition is governed by the binding and unbinding of vortex-antivortex pairs~\cite{berezinskii1971,kosterlitz1973}. Because this transition is topological in character and finite-size effects are pronounced, extracting physically meaningful size-dependent pseudo-critical temperatures from numerical data remains an important problem in computational statistical physics~\cite{nelson1977,hasenbusch2005}.

In recent years, machine learning has become an important tool in statistical physics, especially for identifying transition-related features directly from many-body configurations~\cite{wang2016,carrasquilla2017,vannieuwenburg2017}. Related developments also include feature-engineered approaches to topological structure recognition, applications to strongly correlated systems, and critical examinations of unsupervised phase-discovery pipelines~\cite{zhang2017,chng2017,hu2017}. Supervised and confusion-based approaches have shown that data-driven models can detect signatures of phase-structure changes even when prior information is limited~\cite{carrasquilla2017,vannieuwenburg2017}. For the two-dimensional classical XY model, previous studies have shown that neural networks can detect signals associated with the BKT transition~\cite{beach2018,zhang2019,miyajima2023}, while more recent generative-model approaches have explored thermodynamic and structural information in the same system~\cite{zhang2025}.

The central issue, however, is not merely whether classification can be achieved, but how standard neural-network outputs can be turned into quantitatively testable observables in finite systems. This question is particularly relevant for the 2D XY model, where strong finite-size effects make it desirable to identify a sequence of pseudo-critical temperatures $T^*(L)$ associated with different lattice sizes rather than to rely solely on phase labels. Questions related to input representation, nonlocal correlations, and scale dependence have also been emphasized in the literature~\cite{beach2018,miyajima2023,zhang2020}. More broadly, both supervised and unsupervised studies suggest that, for scientific applications, model outputs should be connected to physically interpretable observables rather than treated merely as classification scores~\cite{zhang2020,carleo2019,greitemann2019,casert2019,rodriguez2019,scheurer2020,yu2021,long2023}.

Motivated by this gap, we develop a framework that extracts a size-dependent sequence of pseudo-critical temperatures $T^*(L)$ from standard classifier outputs in the 2D XY model. The network is trained only on samples from the low-temperature quasi-ordered phase (hereafter, Quasi-ordered Phase) and the high-temperature disordered phase (hereafter, Disordered Phase), while $T^*(L)$ is defined from the 50\% crossing of the corresponding probability curves across the finite-size transition region (hereafter, Critical Region). We then assess this sequence through bootstrap uncertainty, BKT-motivated finite-size scaling, and comparison with conventional thermodynamic indicators.

\section{Model and Methods}\label{sec:methods}

\subsection{Two-Dimensional Classical XY Model}

We consider the classical XY model on a two-dimensional square lattice with periodic boundary conditions. At each lattice site $i$, the spin is a planar unit vector
\begin{equation}
\mathbf{s}_i=(\cos\theta_i,\sin\theta_i),
\end{equation}
where $\theta_i\in[0,2\pi)$ denotes the spin angle. The Hamiltonian is
\begin{equation}
H=-J\sum_{\langle ij\rangle}\cos(\theta_i-\theta_j),
\end{equation}
where $\langle ij\rangle$ denotes nearest-neighbor pairs and $J>0$ is the ferromagnetic coupling constant, favoring local spin alignment. We study lattices with linear sizes $L=16,32,64,128,$ and $256$, corresponding to a total number of sites
\begin{equation}
N=L^2.
\end{equation}
Throughout this paper, we use dimensionless units by setting the ferromagnetic coupling $J=1$ and the Boltzmann constant $k_B=1$. Accordingly, temperature is measured in units of $J/k_B$ and is written simply as the dimensionless temperature $T$.

This model is used here as a source of equilibrium spin configurations from which we extract a size-dependent sequence of pseudo-critical temperatures $T^*(L)$. The chosen lattice sizes allow us to examine how the network-defined temperature scale evolves with system size over a broad finite-size range.

\subsection{Sampling Strategy and Thermodynamic Observables}

To improve sampling efficiency for a continuous-spin system, we employ an embedded cluster-update scheme for the XY model, combining the Swendsen-Wang multi-cluster idea~\cite{swendsen1987} with the random-axis embedding strategy introduced by Wolff for continuous-spin systems~\cite{wolff1989}. In each update, a random projection direction is chosen and the planar spins are projected onto the corresponding local axis, thereby generating Ising-like sign variables. Cluster updates are then carried out in the embedded discrete representation and mapped back to the original XY spin space. Compared with purely local updates, this procedure substantially reduces critical slowing down and is well suited to the generation of large configuration datasets.

In the present implementation, one Monte Carlo step (MCS) is defined operationally as one full embedded-cluster update cycle over the lattice. For the configuration data used in neural-network training and testing, we adopt a parallel independent simulation strategy. At each temperature, 100 independent simulation tasks are launched with different random seeds and random initial conditions. Each task is evolved for 2000 MCS, after which one equilibrium configuration is stored. Because each sample is taken from an independently initialized chain, the resulting dataset is designed to suppress inter-sample autocorrelation. We use 2000 MCS as a uniform dataset-generation setting across all system sizes and temperature ranges considered in this work.

For comparison with the network-defined temperature scale, we also compute conventional thermodynamic observables from independent high-statistics simulations. These include the average energy per site
\begin{equation}
e=\frac{\langle H\rangle}{N},
\end{equation}
the finite-size magnetization
\begin{equation}
m=\frac{1}{N}\left\langle \left|\sum_i e^{i\theta_i}\right| \right\rangle,
\end{equation}
the specific heat
\begin{equation}
C=\frac{\langle H^2\rangle-\langle H\rangle^2}{NT^2},
\end{equation}
and the magnetic susceptibility
\begin{equation}
\chi=\frac{\langle M^2\rangle-\langle M\rangle^2}{NT}, \qquad M=\left|\sum_i e^{i\theta_i}\right|.
\end{equation}
Here $e$ measures the mean internal-energy density, $m$ characterizes the magnitude of the net in-plane spin polarization in a finite system, $C$ quantifies energy fluctuations, and $\chi$ measures fluctuations of the total magnetization. Angle brackets $\langle\cdots\rangle$ denote Monte Carlo averaging. We further define
\begin{equation}
T_{\chi}^{\mathrm{peak}}(L)=\arg\max_T \chi(T,L),
\end{equation}
which serves as a conventional finite-size reference temperature for comparison with the network-extracted pseudo-critical temperature $T^*(L)$.

At each temperature point used for thermodynamic analysis, we perform 20,000 MCS for thermalization followed by 2,000 MCS for measurement averaging. These thermodynamic observables are used only for physical consistency checks and do not participate in neural-network training or in the bootstrap extraction of $T^*(L)$.

\subsection{Configuration Representation and Dataset Construction}

Each equilibrium configuration is first stored as an $L\times L$ array of spin angles $\theta_{ij}\in[0,2\pi)$. To accommodate convolutional neural-network input in a simple and uniform manner, we map the angle field to a single-channel grayscale image according to
\begin{equation}
I_{ij}=\frac{\theta_{ij}\bmod 2\pi}{2\pi},
\end{equation}
where $I_{ij}\in[0,1)$ is the normalized grayscale intensity at lattice site $(i,j)$. The modulo operation preserves angular periodicity, while the linear rescaling maps the angular variable into the grayscale interval used by the classifier.

This encoding is computationally convenient but not symmetry-complete, since it does not preserve the $O(2)$ vector structure as explicitly as the two-channel representation $(\cos\theta,\sin\theta)$. We adopt it here to test whether standard classifier outputs, under a simple and uniform preprocessing pipeline, can still yield physically interpretable pseudo-critical temperatures. Therefore, the conclusions of this work concern the extraction protocol rather than the optimal symmetry-preserving representation.

For each system size, the training set contains only samples drawn well outside the Critical Region. Samples from the Quasi-ordered Phase are taken from the low-temperature interval $T\in[0.001,0.1]$, whereas samples from the Disordered Phase are taken from the high-temperature interval $T\in[3.9,4.0]$. Each interval contains 100 temperature points, and each temperature point contributes 100 independently generated configurations, resulting in 20,000 training samples per system size. The test set covers a size-dependent temperature window centered on the Critical Region, with 60 temperature points and 150 independent samples per temperature, giving 9,000 test samples per system size. The test-temperature windows for all system sizes are listed in Table~S1 of the Supplementary Material.

\begin{figure}[pos=htbp]
\centering
\includegraphics[width=0.95\columnwidth,height=0.7\textheight,keepaspectratio]{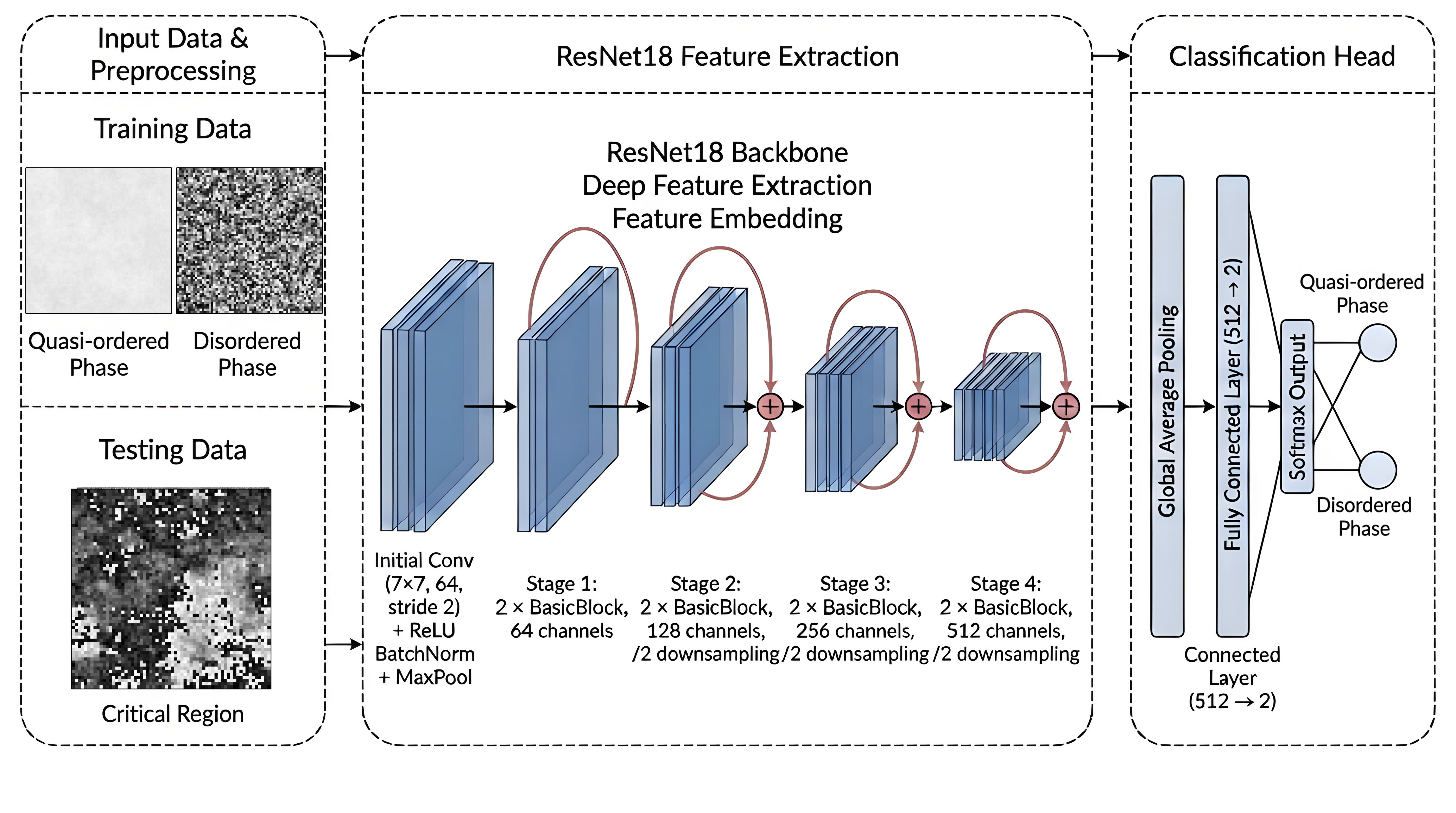}
\caption{ResNet18-based framework for extracting size-dependent temperature scales in the 2D XY model. Monte Carlo configurations are generated from the Quasi-ordered Phase, the Disordered Phase, and the intervening Critical Region. The network is trained using only samples from the Quasi-ordered Phase and the Disordered Phase, and the pseudo-critical temperature $T^*(L)$ is determined from the 50\% crossing of the corresponding probability curves in the Critical Region.}\label{fig:flowchart}
\end{figure}

\subsection{Network Architecture and Training Configuration}

We use a standard ResNet18~\cite{he2016} as the classifier for all system sizes so that the methodological focus remains on the inference protocol rather than on architecture-specific design. This baseline choice also improves reproducibility and portability across lattice models.

Because the input is a single-channel grayscale image, we modify the first convolutional layer of ResNet18 to accept one input channel directly, instead of replicating the grayscale image into three channels. All model parameters are trained from scratch without pretrained weights. For each system size, we train an independent network using only configurations from that size, rather than mixing data from different lattice sizes into a single model.

During training, the data are randomly split into training and validation subsets in an 8:2 ratio. We use cross-entropy loss with equal class weights, the Adam optimizer~\cite{kingma2015}, an initial learning rate of 0.01, and a weight-decay coefficient of $5\times10^{-4}$. The learning-rate schedule follows CosineAnnealingWarmRestarts~\cite{loshchilov2017} with $T_0=5$, $T_{\mathrm{mult}}=2$, and minimum learning rate $\eta_{\min}=5\times10^{-6}$. The batch size is 32 and the total number of training epochs is 20.

To improve robustness while avoiding large distortions of the underlying physical configurations, we apply a conservative augmentation pipeline during training. The geometric transformations include cropping, horizontal and vertical flips, small-angle rotations, and mild affine perturbations. In addition, we apply limited intensity adjustment and random erasing as regularization-style perturbations; random erasing is a standard augmentation strategy for improving the robustness of convolutional networks~\cite{zhong2020}. At inference time, we further adopt test-time augmentation (TTA), in which each configuration is evaluated not only in its original form but also under a small set of additional transformed views, and the final class probability is obtained by averaging the corresponding predictions~\cite{shanmugam2020}. Compared with the training-time augmentation, the TTA transforms are deliberately more conservative in rotation, translation, and scale so that the reported probabilities remain closely tied to the original physical configuration.

The overall training-and-inference workflow is summarized in Fig.~\ref{fig:flowchart}. For each lattice size, Monte Carlo configurations sampled from the low-temperature quasi-ordered regime and the high-temperature disordered regime are first converted into single-channel grayscale inputs and then used to train an independent ResNet18 classifier. Configurations from the intermediate Critical Region are not used for supervised training; instead, they are passed through the trained network at inference to obtain class probabilities, which are then averaged over samples and test-time augmentations. The size-dependent pseudo-critical temperature $T^*(L)$ is finally extracted from the 50\% crossing of the quasi-ordered and disordered probability curves.

\subsection{50\% Probability Crossing Temperature and Finite-Size Scaling}

For the test set covering the Critical Region, the trained network assigns to each input configuration two probabilities: the probability of belonging to the Quasi-ordered Phase and the probability of belonging to the Disordered Phase. After averaging over samples at the same temperature, we denote these quantities by $P_{\mathrm{QO}}(T,L)$ and $P_{\mathrm{D}}(T,L)$, respectively, with
\begin{equation}
P_{\mathrm{QO}}(T,L)+P_{\mathrm{D}}(T,L)=1.
\end{equation}
As the temperature increases across the Critical Region, $P_{\mathrm{QO}}(T,L)$ decreases while $P_{\mathrm{D}}(T,L)$ increases. We define the pseudo-critical temperature $T^*(L)$ as the temperature at which the two probabilities become equal,
\begin{equation}
P_{\mathrm{QO}}(T^*(L),L)=P_{\mathrm{D}}(T^*(L),L)=0.5.
\end{equation}
This 50\% crossing criterion provides a simple operational definition of the size-dependent pseudo-critical temperature and identifies the center of the classifier response across the Critical Region.

To estimate the statistical uncertainty of $T^*(L)$, we use bootstrap resampling~\cite{efron1993}. For each lattice size and temperature, the configuration-level prediction probabilities are resampled with replacement to generate bootstrap replicas of the averaged probability curves. The crossing temperature is then determined for each bootstrap replica by interpolation between neighboring temperature points. From the resulting bootstrap distribution, we obtain the mean pseudo-critical temperature $\bar{T}^*(L)$ and its standard deviation $\sigma_{T^*}(L)$, which we use as the reported estimate and uncertainty. The reported uncertainty reflects finite-sample variability at fixed model parameters and temperature grid, and does not include additional variation from training randomness, network initialization, or data splitting.

To examine whether the extracted pseudo-critical temperatures are compatible with known finite-size physics of the 2D XY model, we analyze $T^*(L)$ using the standard BKT-motivated finite-size scaling form
\begin{equation}
T^*(L)=T_{\mathrm{BKT}}+\frac{a}{[\ln L+b\ln(\ln L)]^2},
\end{equation}
where $T_{\mathrm{BKT}}$ is the thermodynamic-limit transition temperature and $a$ and $b$ are fitting parameters. A brief theoretical background for the BKT transition temperature and the scaling form used here is provided in the Supplementary Material. We also introduce the rescaled variable
\begin{equation}\label{eq:bkt_rescaled}
x=[T-T_{\mathrm{BKT}}][\ln L+b\ln(\ln L)]^2,
\end{equation}
which is used to compare the temperature dependence of the probability curves for different lattice sizes on a common finite-size scaling axis.

\section{Results and Analysis}\label{sec:results}

\subsection{Thermodynamic Observables and the Critical Region}

In this section, we first identify the finite-size temperature interval highlighted by conventional thermodynamic observables, and then examine whether the pseudo-critical temperatures extracted from the classifier response follow the same finite-size physics.

\begin{figure}[pos=htbp]
\centering
\includegraphics[width=0.95\columnwidth,height=0.7\textheight,keepaspectratio]{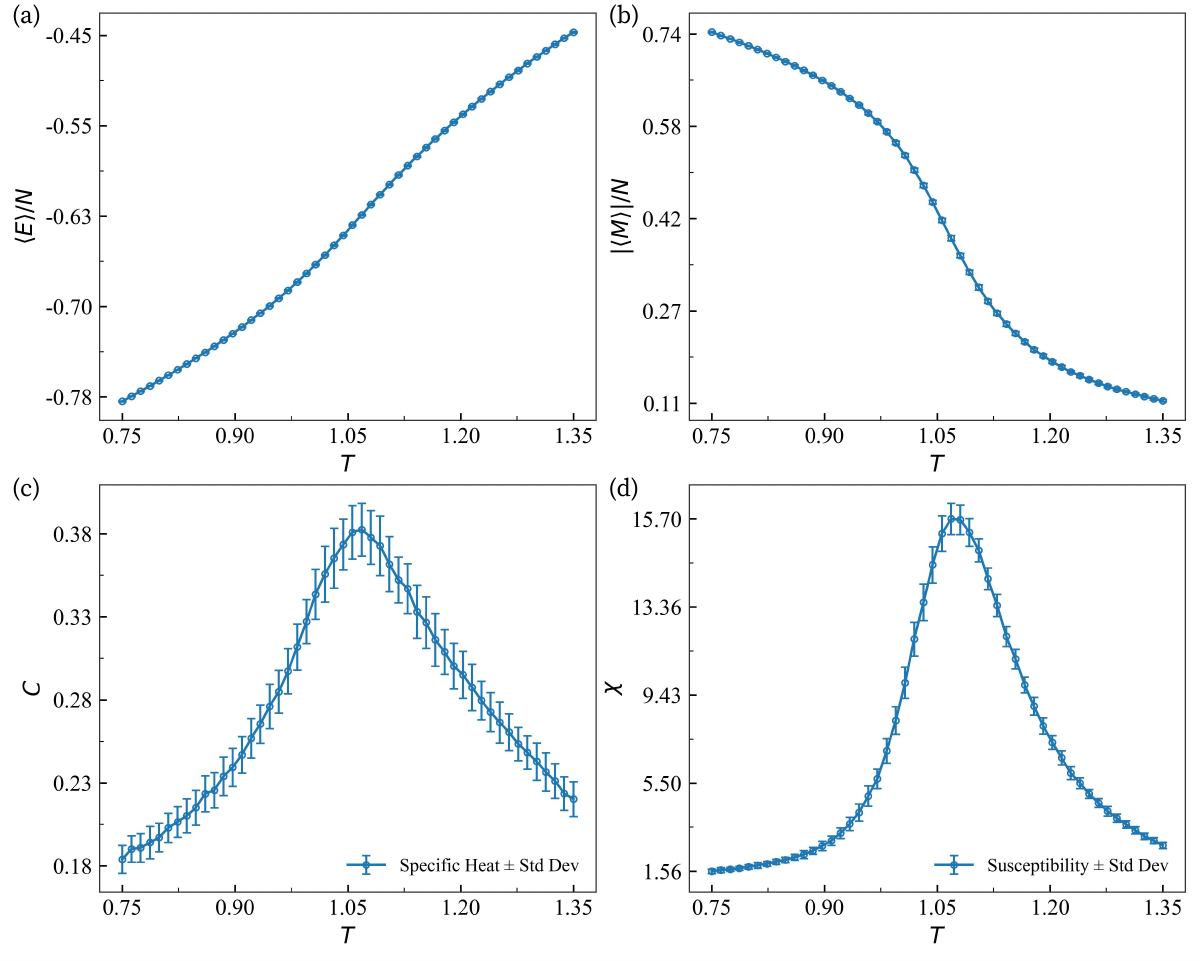}
\caption{Thermodynamic quantities for the $32\times32$ system as functions of temperature. (a) Average energy per site. (b) Finite-size magnetization. (c) Specific heat. (d) Susceptibility.}\label{fig:thermodynamics}
\end{figure}

Figure~\ref{fig:thermodynamics} presents several representative thermodynamic observables for the $32\times32$ system, including the energy per site, finite-size magnetization, specific heat, and susceptibility. As the temperature increases, the energy changes smoothly and the finite-size magnetization decreases continuously, while the specific heat and susceptibility exhibit pronounced peak-like structures within a common temperature interval. For a finite lattice, this interval identifies the temperature range in which thermodynamic fluctuations are strongest and the system undergoes its most rapid finite-size change. In the present work, we label this interval as the Critical Region.

We therefore use this fluctuation-dominated interval as the Critical Region, i.e., the physically motivated finite-size temperature window against which the classifier-defined temperature $T^*(L)$ is interpreted. The $32\times32$ system is shown here as a representative example; additional small-system thermodynamic curves are provided in Figs.~S1--S3 of the Supplementary Material, and the corresponding simulation parameters are summarized in Table~S2.

\subsection{Neural-network probability curves and finite-size scaling of the extracted pseudo-critical temperature}

\begin{figure}[pos=htbp]
\centering
\includegraphics[width=0.95\columnwidth,height=0.7\textheight,keepaspectratio]{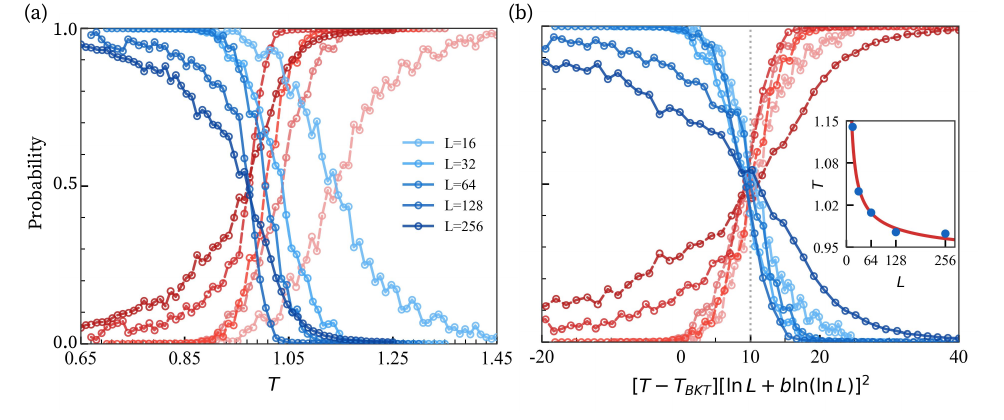}
\caption{Neural-network probability curves and BKT finite-size scaling analysis for the 2D XY model across the studied sizes. (a) Probabilities assigned to the Quasi-ordered Phase and the Disordered Phase as functions of temperature. (b) Probability-curve collapse based on the BKT rescaled variable, with the inset showing the finite-size fit of the 50\% probability crossing temperature $T^*(L)$.}\label{fig:probability}
\end{figure}

Figure~\ref{fig:probability} summarizes the central result of this work. As shown in Fig.~\ref{fig:probability}(a), the probability assigned to the Quasi-ordered Phase decreases with temperature, whereas the probability assigned to the Disordered Phase increases correspondingly. For all studied lattice sizes, the two curves intersect at a well-defined temperature, which we identify as the pseudo-critical temperature $T^*(L)$. As the system size increases, the crossover becomes progressively sharper and the crossing point shifts systematically toward lower temperature, indicating that the network output tracks a size-dependent finite-size temperature scale rather than a fixed classification threshold.

The extracted values of $T^*(L)$, listed in Table~S3 of the Supplementary Material, exhibit a clear finite-size drift. The associated bootstrap uncertainties remain narrow for all studied system sizes, showing that the crossing temperature is statistically stable against finite-sample fluctuations at fixed trained parameters. In this sense, the 50\% probability-crossing criterion provides a robust operational way to convert standard classifier outputs into a characteristic temperature scale. Detailed bootstrap summary statistics are provided in Table~S4 of the Supplementary Material.

To examine whether this extracted temperature scale is physically meaningful, we further analyze the probability curves using the BKT-motivated rescaled variable defined in Eq.~(\ref{eq:bkt_rescaled}). As shown in Fig.~\ref{fig:probability}(b), the curves for different lattice sizes display a partial collapse onto a common trend after rescaling. Although the collapse is not exact, which is expected for finite lattices and a limited range of system sizes, it nevertheless indicates that the classifier response across the Critical Region follows the same finite-size tendency for different $L$.

The inset of Fig.~\ref{fig:probability}(b) shows the finite-size fit of the extracted crossing temperatures $T^*(L)$ using Eq.~(10). The fitted curve captures the logarithmically slow drift of the crossing temperature with increasing system size, consistent with the expected BKT-type finite-size behavior near the transition. The corresponding fitting parameters and residual diagnostics are provided in Tables~S5 and S6 of the Supplementary Material. Taken together, Fig.~\ref{fig:probability}(a), Fig.~\ref{fig:probability}(b), and its inset show that the classifier output can be converted into a statistically stable and physically interpretable sequence of size-dependent pseudo-critical temperatures. This analysis is intended as a finite-size consistency check rather than a precision determination of the thermodynamic-limit transition temperature.

\subsection{Comparison with Susceptibility Peak Temperature}

\begin{figure}[pos=htbp]
\centering
\includegraphics[width=0.95\columnwidth,height=0.7\textheight,keepaspectratio]{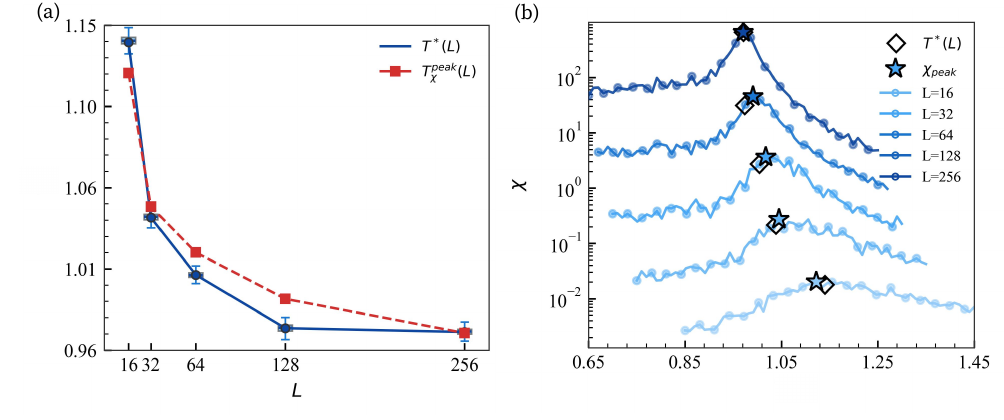}
\caption{Comparison of size-dependent pseudo-critical temperatures $T^*(L)$ extracted from the classifier response with susceptibility-peak temperatures $T_{\chi}^{\mathrm{peak}}(L)$. (a) $T^*(L)$ and $T_{\chi}^{\mathrm{peak}}(L)$ as functions of system size. (b) Susceptibility as a function of temperature for the studied system sizes, with peak positions and 50\% probability crossing temperatures marked.}\label{fig:susceptibility}
\end{figure}

As an additional physical consistency check, we compare the pseudo-critical temperature $T^*(L)$, extracted from the classifier response across the Critical Region, with the susceptibility-peak temperature $T_{\chi}^{\mathrm{peak}}(L)$. Figure~\ref{fig:susceptibility}(a) shows that both temperature scales move toward lower temperature as the lattice size increases, while Fig.~\ref{fig:susceptibility}(b) shows that they remain within the same finite-size temperature window. The detailed comparison is summarized in Table~S3 of the Supplementary Material.

This agreement is physically significant because both $T^*(L)$ and $T_{\chi}^{\mathrm{peak}}(L)$ probe the same finite-size fluctuation window, although they are defined from different observables and therefore need not coincide at finite $L$. For smaller lattices, finite-size rounding and observable-dependent response lead to a visible offset between the two temperature scales. As the system size increases, however, this offset becomes overall smaller and is nearly negligible for the largest lattice studied, indicating that the two definitions are converging toward the same thermodynamic transition scale. This trend further supports the interpretation of $T^*(L)$ as a physically meaningful size-dependent pseudo-critical temperature. At the same time, we emphasize that the present results establish convergence in a finite-size sense, rather than a precision proof that the thermodynamic-limit transition temperature has already been exactly determined by the classifier.

\section{Discussion and Conclusion}\label{sec:discussion}

Our results show that standard classifier probabilities can be converted into a statistically stable and physically interpretable sequence of size-dependent pseudo-critical temperatures in the 2D XY model. The extracted temperature scale follows a BKT-consistent finite-size drift and remains closely connected to the susceptibility-defined fluctuation window, indicating that the network response is not merely algorithmic but reflects thermodynamically meaningful finite-size reorganization of configuration statistics.

At the same time, the present analysis should be interpreted at the level of finite-size inference rather than as a precision determination of the thermodynamic-limit BKT transition temperature. In this sense, the main contribution of this work is methodological: it provides a practical and reproducible framework for turning standard neural-network outputs into quantitatively testable finite-size observables. Because the procedure relies only on a baseline architecture and a simple preprocessing pipeline, it may also be extended to other statistical systems in which crossover behavior, strong finite-size effects, or topological transition signatures make physically interpretable temperature-scale extraction nontrivial.

Several extensions remain open. A more systematic analysis of thermalization near the Critical Region, especially for larger lattice sizes, would strengthen the statistical control of the extracted temperatures. It would also be useful to compare the present grayscale representation with symmetry-preserving inputs such as $(\cos\theta,\sin\theta)$, in order to separate more clearly the role of the inference protocol from that of the input representation. More broadly, this work helps bridge the gap between machine-learning-based phase recognition and physically grounded finite-size analysis in complex statistical systems.





\printcredits

\section*{Supplementary information}

Supplementary information includes: MC sampling parameters (Table~S1), theoretical background for the BKT transition temperature (Sec.~S2), thermodynamic quantities for additional small systems (Figs.~S1--S3), Monte Carlo simulation parameters for thermodynamic analysis (Table~S2), comparison of network-extracted and susceptibility-peak temperatures (Table~S3), bootstrap statistical results (Table~S4), BKT fitting parameters (Table~S5), residual analysis (Table~S6), validation confusion matrices (Fig.~S4), and zoomed probability curves in the crossing region (Fig.~S5).

\section*{Acknowledgements}

The authors acknowledge the computational resources provided by the institution. Helpful discussions with colleagues are gratefully appreciated.

\section*{Declarations}

\begin{itemize}
\item \textbf{Funding:} This work was supported by the National Natural Science Foundation of China (Grant No.~12135003).
\item \textbf{Conflict of interest:} The authors declare no conflict of interest.
\item \textbf{Ethics approval:} Not applicable
\item \textbf{Data availability:} The GitHub repository for this work contains all code required to reproduce the results, including Monte Carlo data generation, model training, bootstrap analysis, and finite-size scaling fitting. It also provides representative example datasets for illustration of the data format and usage, trained model weights, the extracted $T^*(L)$ values with bootstrap uncertainties, and the processed data used to generate the main figures. The full raw Monte Carlo simulation datasets are not included due to their large size (tens of GB) but are available from the authors upon reasonable request. Documentation describing the data structure and usage is provided in the repository. The repository is accessible at \url{https://github.com/p17853087608-collab/XY-Model}.
\item \textbf{Code availability:} The Python code used in this study is publicly available at \url{https://github.com/p17853087608-collab/XY-Model}.
\end{itemize}

\FloatBarrier
\bibliographystyle{model1-num-names}

\bibliography{my_paper}



\end{document}


\title{\Large\bfseries Supplementary Materials for\\[0.3em]
``Machine-learning extraction of size-dependent temperature scales in the 2D XY model''}

\author{Qingao Fan\textsuperscript{1,2},\quad Xu Li\textsuperscript{1,2,*},\quad Tingting Xue\textsuperscript{1,2,*}}

\date{}

\maketitle

\noindent\textsuperscript{1}Faculty of Science, Kunming University of Science and Technology, Kunming 650500, Yunnan, China\\[0.3em]
\noindent\textsuperscript{2}Yunnan Key Laboratory of Complex Systems and Brain-Inspired Intelligence, Kunming University of Science and Technology, Kunming 650500, Yunnan, China\\[1em]
\noindent\textsuperscript{*}Corresponding authors. E-mail: \href{mailto:lixucn@kust.edu.cn}{lixucn@kust.edu.cn} (X.\,Li); \href{mailto:tingtingxue@kust.edu.cn}{tingtingxue@kust.edu.cn} (T.\,Xue)

\vspace{1.5em}

\noindent\textbf{Abstract.} These Supplementary Materials provide methodological details and additional results supporting the machine-learning extraction of the size-dependent pseudo-critical temperature $T^*(L)$ in the 2D XY model.

\vspace{1.5em}


\section*{S1. Data Generation Protocol}

This section documents the data-generation procedure underlying the neural-network probability curves and the extracted pseudo-critical temperatures described in Secs.~2.3--2.4 and Sec.~3.2 of the main text. Table~S1 specifies the temperature intervals and sampling parameters used for training and testing.

\begin{table}[htbp]
\caption{Monte Carlo data-sampling parameters. The Quasi-ordered Phase and the Disordered Phase use identical parameters across all five system sizes ($L=16$--$256$); the Critical Region uses system-size-specific $T$ ranges.}
\label{tab:S1}
\centering
\begin{tabular}{@{}llcccc@{}}
\toprule
\multirow{2}{*}{System Size $L$} & \multirow{2}{*}{Temperature Region} & \multirow{2}{*}{$T$ Range} & Sample & Samples & Total \\
 &  &  & Points & per Point & Samples \\
\midrule
\multirow{2}{*}{All sizes}
  & Quasi-ordered Phase & $0.001$--$0.1$ & 100 & 100 & 10,000 \\
  & Disordered Phase    & $3.9$--$4.0$   & 100 & 100 & 10,000 \\
\midrule
16  & Critical Region & $0.85$--$1.45$ & 60  & 150 & 9,000 \\
32  & Critical Region & $0.75$--$1.35$ & 60  & 150 & 9,000 \\
64  & Critical Region & $0.70$--$1.30$ & 60  & 150 & 9,000 \\
128 & Critical Region & $0.67$--$1.27$ & 60  & 150 & 9,000 \\
256 & Critical Region & $0.65$--$1.25$ & 60  & 150 & 9,000 \\
\bottomrule
\end{tabular}

\begin{minipage}{\textwidth}
\vspace{0.5em}
\footnotesize
Data generated via parallel Monte Carlo simulations using the embedded cluster-update scheme. For each temperature point: 2000 MCS for equilibration, followed by configuration saving. Each sample represents an independently generated configuration. $T^*(L)$ is obtained from the 50\% crossing of bootstrap-resampled averaged probability curves by interpolation between neighboring sampled temperature points, consistent with Sec.~2.5 of the main text. $T_{\chi}^{\mathrm{peak}}(L)$ is taken as the temperature at which $\chi(T,L)$ reaches its maximum on the sampled temperature grid, consistent with Sec.~2.2 of the main text.
\end{minipage}
\end{table}

\clearpage


\section*{S2. Theoretical Background for the BKT Transition Temperature}

The Berezinskii-Kosterlitz-Thouless (BKT) transition in the two-dimensional XY model is a topological phase transition driven by the binding and unbinding of vortex-antivortex pairs~\cite{nelson1977,hasenbusch2005}. Below the transition, vortices exist only as tightly bound pairs, producing quasi-long-range order characterized by a power-law decay of the spin correlation function. Above the transition, free vortices proliferate, leading to exponential decay of correlations.

A rigorous criterion for the thermodynamic-limit transition temperature $T_{\mathrm{BKT}}$ is provided by the universal jump condition for the superfluid stiffness (helicity modulus)~\cite{nelson1977}:
\begin{equation}
\Upsilon(T_{\mathrm{BKT}}) = \frac{2}{\pi} T_{\mathrm{BKT}}.
\label{eq:S1}
\end{equation}
The helicity modulus itself is not computed in the present work; instead, we adopt the high-precision reference value $T_{\mathrm{BKT}}=0.8929$ obtained by Hasenbusch~\cite{hasenbusch2005} from large-scale Monte Carlo simulations using the universal jump condition. This value is fixed in the finite-size scaling analysis of the main text.

The finite-size scaling form used in Eq.~(10) of the main text is
\begin{equation}
T^*(L) = T_{\mathrm{BKT}} + \frac{a}{[\ln L + b\ln(\ln L)]^2},
\label{eq:S2}
\end{equation}
motivated by the BKT theory prediction that, for a finite lattice of linear size $L$, the pseudo-critical temperature approaches $T_{\mathrm{BKT}}$ logarithmically slowly with increasing $L$. The parameter $b$ captures sub-leading logarithmic corrections arising from the marginally irrelevant operator at the BKT fixed point. This form has been widely used in previous finite-size studies of the 2D XY model~\cite{hasenbusch2005,ueda2021}. The numerical fit of this form to the extracted $T^*(L)$ and the resulting best-fit parameters are presented in Table~S5 of Sec.~S6.

\clearpage


\section*{S3. Thermodynamic Observables as Conventional Finite-Size Reference Quantities}

In addition to the machine-learning results, we evaluate several conventional thermodynamic observables from independent Monte Carlo samples. These quantities are used only for physical consistency checks and finite-size comparison, and do not enter the neural-network training or the bootstrap extraction of $T^*(L)$. In particular, the susceptibility peak temperature provides a conventional finite-size reference scale for comparison with the network-extracted pseudo-critical temperature discussed in Sec.~3.1 and Sec.~3.3 of the main text. The comparison data are further detailed in Table~S3 of Sec.~S5. The Monte Carlo simulation parameters used for the thermodynamic analysis are listed in Table~S2 of Sec.~S4. The definitions of these observables are given in Sec.~2.2 of the main text.

\begin{figure}[htbp]
\centering
\includegraphics[width=0.9\textwidth]{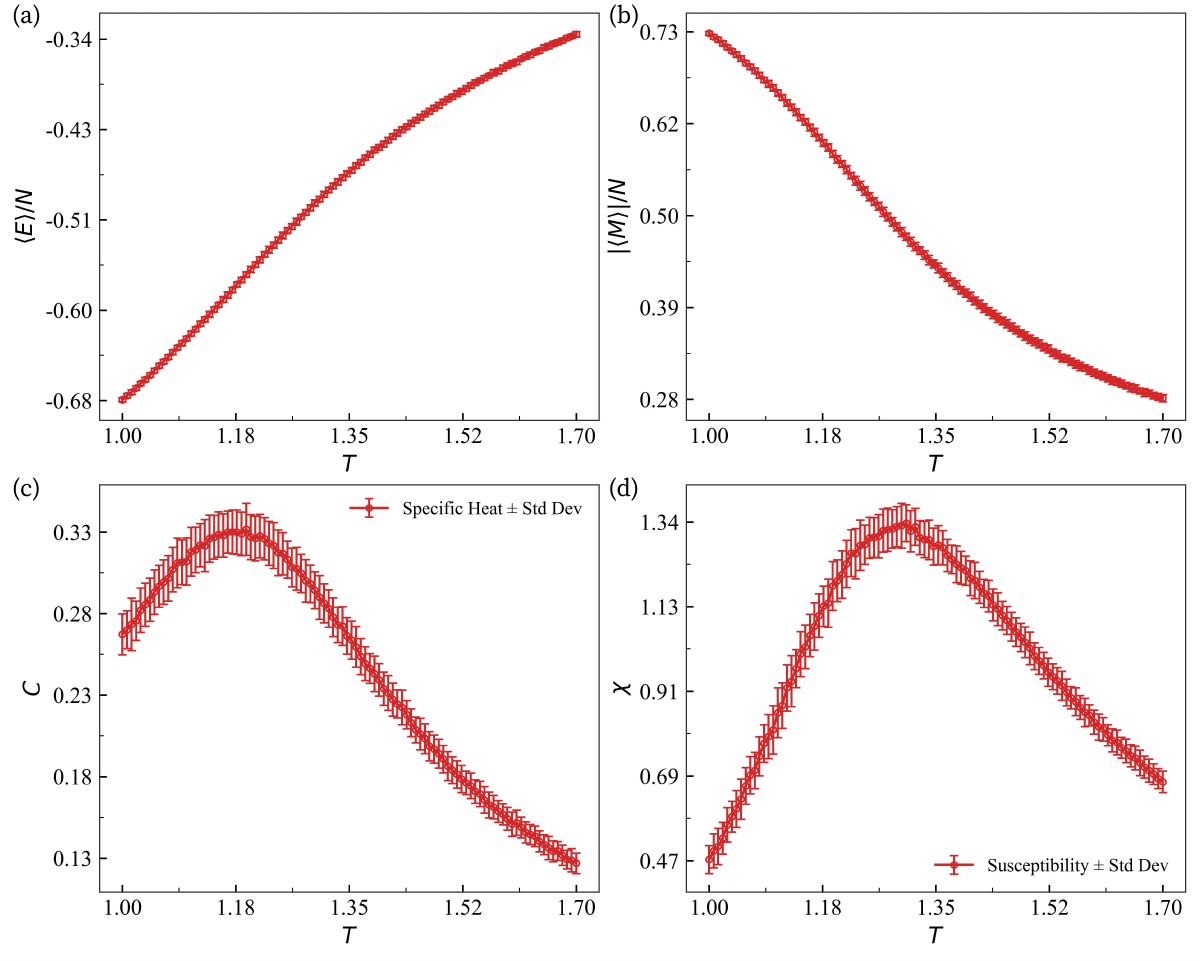}
\caption{Thermodynamic observables for the $8\times8$ system as functions of temperature: (a) average energy per site, (b) finite-size magnetization, (c) specific heat, and (d) susceptibility.}
\label{fig:S1}
\end{figure}

\begin{figure}[htbp]
\centering
\includegraphics[width=0.9\textwidth]{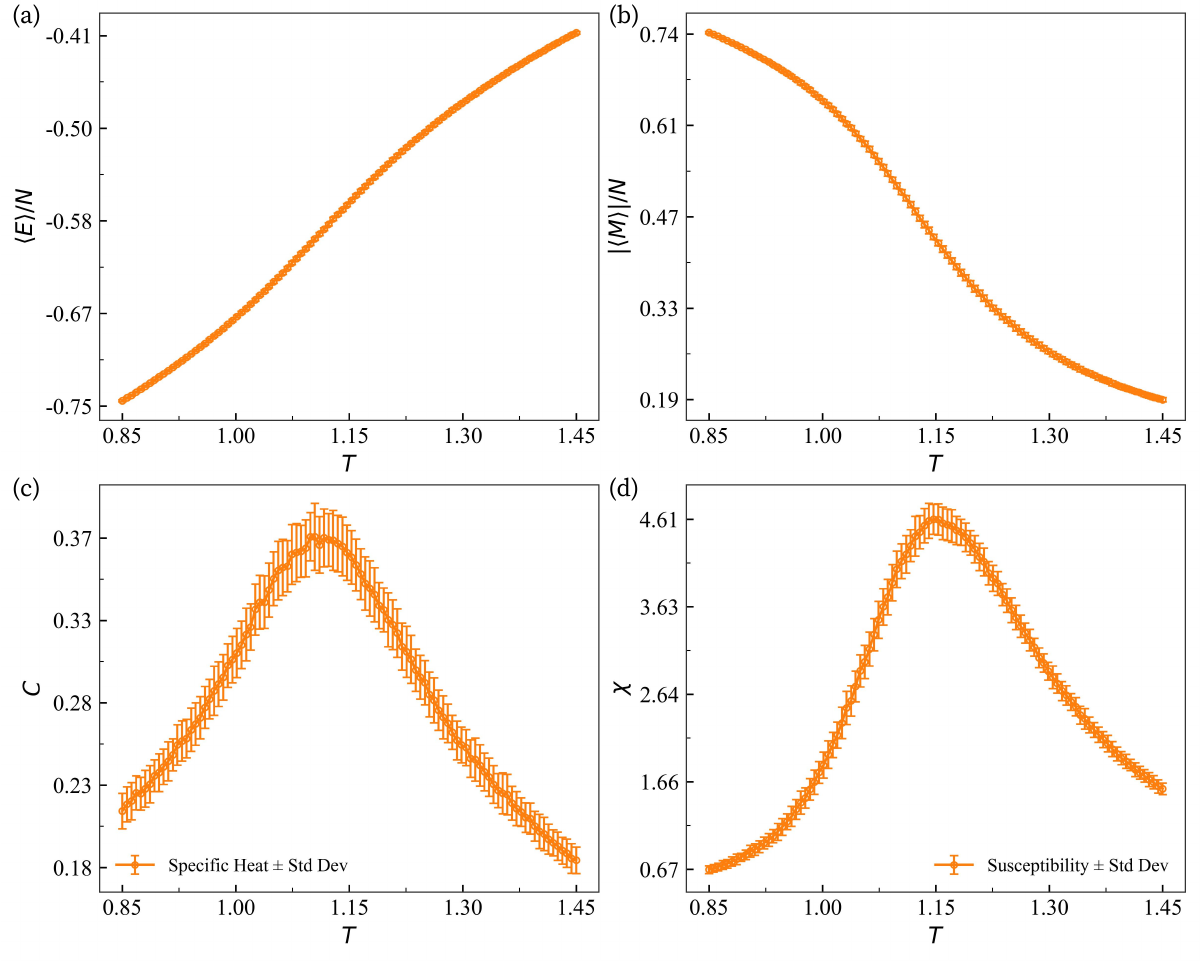}
\caption{Thermodynamic observables for the $16\times16$ system as functions of temperature: (a) average energy per site, (b) finite-size magnetization, (c) specific heat, and (d) susceptibility.}
\label{fig:S2}
\end{figure}

\begin{figure}[htbp]
\centering
\includegraphics[width=0.9\textwidth]{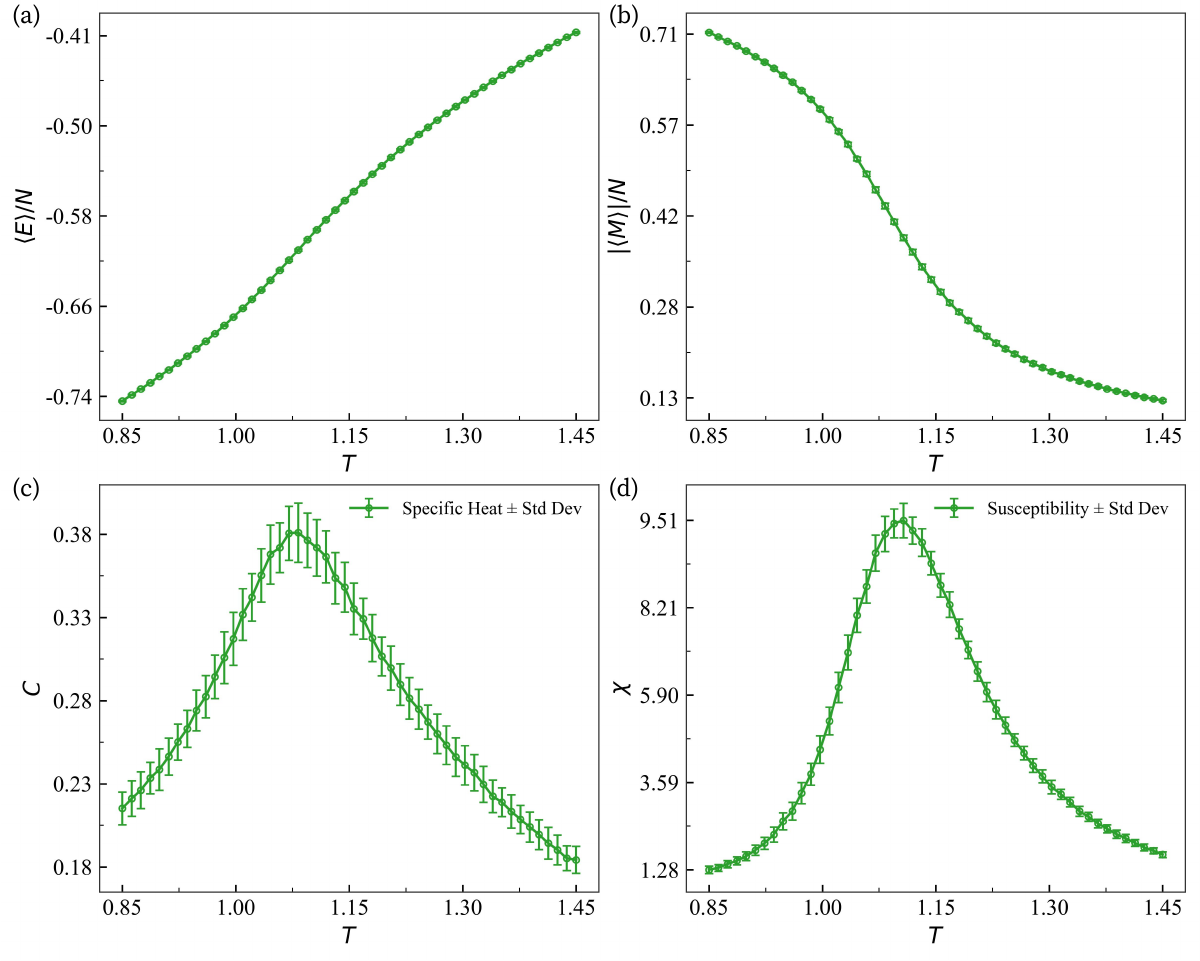}
\caption{Thermodynamic observables for the $24\times24$ system as functions of temperature: (a) average energy per site, (b) finite-size magnetization, (c) specific heat, and (d) susceptibility.}
\label{fig:S3}
\end{figure}

\clearpage


\section*{S4. Monte Carlo Simulation Parameters for Thermodynamic Analysis}

Table~S2 documents the simulation parameters used for the conventional thermodynamic analysis presented in Sec.~S3. These data were generated independently from the neural-network training data, using the same embedded cluster-update framework described in Sec.~2.2 of the main text.

\vspace{6pt}
\begin{table}[htbp]
\caption{Monte Carlo simulation parameters for thermodynamic analysis using the embedded cluster-update scheme. Data used for generating temperature-dependent thermodynamic quantities ($\langle E \rangle/N$, $|\langle M \rangle|/N$, $C$, and $\chi$) shown in Figs.~S1--S3.}
\label{tab:S2}
\centering
\begin{tabular}{@{}lrrrr@{}}
\toprule
Parameter & $L=8$ & $L=16$ & $L=24$ & $L=32$ \\
\midrule
$T$ Range              & 1.00--1.70 & 0.85--1.45 & 0.85--1.45 & 0.75--1.35 \\
Sample Points     & 100 & 100 & 50 & 50 \\
Samples per Point & 100 & 100 & 50 & 50 \\
\bottomrule
\end{tabular}

\begin{minipage}{\textwidth}
\vspace{0.5em}
\footnotesize
Each independent run performs a complete temperature scan from low to high temperature. For each temperature point: (1)\,20\,000 MC steps for equilibration using the embedded cluster-update scheme; (2)\,2\,000 MC steps for thermodynamic measurement averaging. Periodic boundary conditions applied.
\end{minipage}
\end{table}

\vspace{2em}


\section*{S5. Comparison of Network-Extracted and Susceptibility-Peak Temperatures}

Table~S3 summarizes the size-dependent pseudo-critical temperatures $T^*(L)$ extracted from the classifier response together with the susceptibility-peak temperatures $T_{\chi}^{\mathrm{peak}}(L)$. The temperature difference $\Delta T = T^*(L) - T_{\chi}^{\mathrm{peak}}(L)$ remains small across all studied lattice sizes, consistent with the finite-size convergence discussed in Sec.~3.3 of the main text.

\begin{table}[htbp]
\caption{Size-dependent pseudo-critical temperatures $T^*(L)$ extracted from the classifier response and comparison with susceptibility-peak temperatures $T_{\chi}^{\mathrm{peak}}(L)$. $T^*(L)$ values obtained via bootstrap analysis ($N_{\mathrm{boot}} = 1000$). $T_{\chi}^{\mathrm{peak}}(L)$ obtained by locating the discrete maximum of susceptibility on the temperature grid. Here $\Delta T = T^*(L) - T_{\chi}^{\mathrm{peak}}(L)$.}
\label{tab:S3}
\centering
\begin{tabular}{@{}crrr@{}}
\toprule
$L$ & $T^*(L)$ & $T_{\chi}^{\mathrm{peak}}$ & $\Delta T$ \\
\midrule
16  & $1.1418 \pm 0.0056$ & 1.1228 & +0.0190 \\
32  & $1.0390 \pm 0.0031$ & 1.0448 & $-0.0058$ \\
64  & $1.0050 \pm 0.0018$ & 1.0181 & $-0.0131$ \\
128 & $0.9742 \pm 0.0025$ & 0.9910 & $-0.0168$ \\
256 & $0.9720 \pm 0.0029$ & 0.9711 & +0.0009 \\
\bottomrule
\end{tabular}
\end{table}

\clearpage


\section*{S6. Statistical Robustness of the Extracted Pseudo-Critical Temperatures}

This section provides additional statistical details for the pseudo-critical temperatures $T^*(L)$ reported in the main text. Table~S4 summarizes the bootstrap uncertainty, supplementing Sec.~2.5 and Sec.~3.2 of the main text.

\vspace{6pt}
\begin{table}[htbp]
\caption{Bootstrap summary statistics for $T^*(L)$ based on $N_{\mathrm{boot}}=1000$ resamples. SD: standard deviation; CV: coefficient of variation ($\mathrm{CV}=\mathrm{SD}/\bar{T}^*\times100\%$); $Q_1$/$Q_3$: first/third quartiles of the bootstrap distribution.}
\label{tab:S4}
\centering
\begin{tabular}{@{}crrrrr@{}}
\toprule
$L$ & SD & CV (\%) & $Q_1$ & $Q_3$ & IQR \\
\midrule
16  & 0.00561 & 0.4918 & 1.1397 & 1.1436 & 0.0039 \\
32  & 0.00309 & 0.2972 & 1.0372 & 1.0405 & 0.0032 \\
64  & 0.00182 & 0.1807 & 1.0037 & 1.0063 & 0.0026 \\
128 & 0.00253 & 0.2599 & 0.9724 & 0.9758 & 0.0034 \\
256 & 0.00292 & 0.3006 & 0.9706 & 0.9734 & 0.0028 \\
\bottomrule
\end{tabular}
\end{table}

\vspace{1em}

\begin{table}[htbp]
\caption{Parameters of the BKT-motivated finite-size fit for $T^*(L)$, with $T_{\mathrm{BKT}}=0.8929$ fixed to the reference value of Hasenbusch~\cite{hasenbusch2005}. The fit uses all five system sizes $L = \{16, 32, 64, 128, 256\}$.}
\label{tab:S5}
\centering
\begin{tabular}{@{}lr@{}}
\toprule
Parameter & Value \\
\midrule
$T_{\mathrm{BKT}}$ & 0.8929 (fixed) \\
$a$ & 10.3001 \\
$b$ & 3.8169 \\
$\chi^2/\mathrm{df}$ & 9.0328 \\
\bottomrule
\end{tabular}

\vspace{1em}
\noindent\textbf{Fitting equation:}
\begin{equation*}
T^*(L) = 0.8929 + \dfrac{10.3001}{[\ln L + 3.8169 \ln(\ln L)]^2}
\end{equation*}
\end{table}

\vspace{1em}

\begin{table}[htbp]
\caption{Residual analysis comparing observed $T^*(L)$ values with theoretical predictions from the BKT finite-size fit. $T^{*\,\mathrm{obs}}$ denotes the bootstrap mean, and $T^{*\,\mathrm{theo}}$ is computed from Eq.~\eqref{eq:S2}. RMSE: root mean square error; MAPE: mean absolute percentage error.}
\label{tab:S6}
\centering
\begin{tabular}{@{}crrrrr@{}}
\toprule
$L$ & $T^{*\,\mathrm{obs}}$ & $T^{*\,\mathrm{theo}}$ & Residual & $|$Residual$|$ & $|$Residual$|/\%$ \\
\midrule
16  & 1.1418 & 1.1248 & $+0.0170$ & 0.0170 & 1.49 \\
32  & 1.0390 & 1.0457 & $-0.0067$ & 0.0067 & 0.65 \\
64  & 1.0050 & 1.0047 & $+0.0003$ & 0.0003 & 0.03 \\
128 & 0.9742 & 0.9799 & $-0.0057$ & 0.0057 & 0.59 \\
256 & 0.9720 & 0.9634 & $+0.0086$ & 0.0086 & 0.88 \\
\midrule
\multicolumn{6}{c}{Residual Statistics} \\
\cmidrule(lr){2-6}
& Mean $|$Res.$|$ & Max $|$Res.$|$ & RMSE & MAPE (\%) & ~ \\
& 0.0077 & 0.0170 & 0.0094 & 0.73 & ~ \\
\bottomrule
\end{tabular}
\end{table}

\clearpage


\section*{S7. Classifier Validation}

This section validates the phase-classification performance of the neural network. Fig.~S4 shows the validation confusion matrices for all five system sizes. The near-perfect classification consistency indicates that the subsequent extraction of $T^*(L)$ is not limited by misclassification within the labeled training regime.

\begin{figure}[htbp]
\centering
\includegraphics[width=0.9\textwidth]{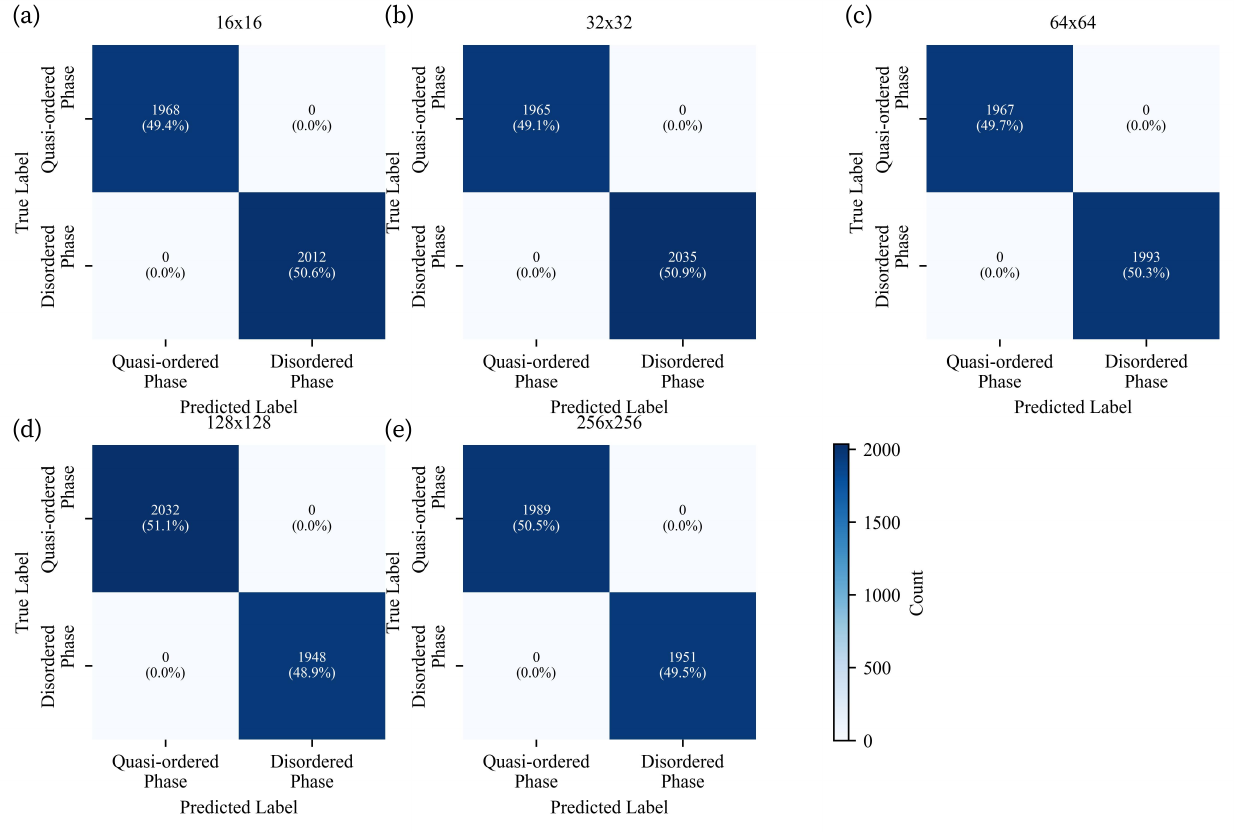}
\caption{Validation confusion matrices for $L=16$, $32$, $64$, $128$, and $256$. Each matrix corresponds to the validation set constructed from the Quasi-ordered Phase and the Disordered Phase.}
\label{fig:S4}
\end{figure}

\clearpage


\section*{S8. Probability-Curve Crossing Analysis for Defining $T^*(L)$}

To make the extraction of $T^*(L)$ visually transparent, this section presents a zoomed view of the probability curves in the Critical Region. The 50\% crossing of the relevant probability is used to define $T^*(L)$ for each system size, and the systematic shift of the crossing toward lower temperatures with increasing $L$ can be directly observed. This figure supplements the main-text discussion in Sec.~3.2.

\begin{figure}[htbp]
\centering
\includegraphics[width=0.7\textwidth]{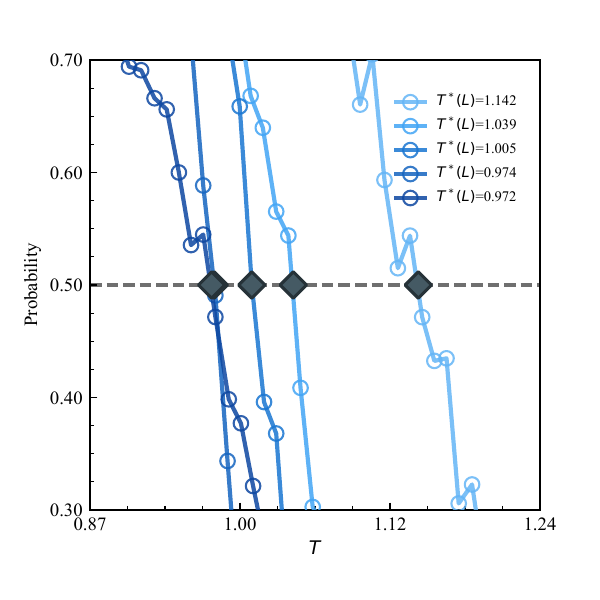}
\caption{Zoomed view of the Critical Region probability curves used to extract $T^*(L)$. The marked 50\% crossings define $T^*(L)$ for each system size. The systematic leftward shift with increasing $L$ illustrates the finite-size drift discussed in Sec.~3.2 of the main text.}
\label{fig:S5}
\end{figure}

\bibliographystyle{unsrtnat}
\bibliography{my_paper}